\documentstyle[12pt]{article}
\begin{document}
\title{\bf Too Soon for Doom Gloom?
\thanks{Alberta-Thy-17-94, gr-qc/9407002.}}
\author{
Tom\'{a}\v{s} Kopf\thanks{Present address:  Mathematical Institute,
Silesian University in Opava, Na Rybnicku 1,
746 01 Opava, Czech Republic; Internet address:
tomas.kopf@math.slu.cz}
\\
Pavel Krtou\v{s}\thanks{Present address:  Institute of Theoretical Physics, Faculty of Mathematics and Physics, Charles University in Prague, V Hole\v{s}ovi\v{c}k\'{a}ch 2, CZ-180 00, Praha 8, Czech Republic; Internet address:
Pavel.Krtous@utf.mff.cuni.cz}
\\ and\\
Don N. Page
\thanks{Internet address:
profdonpage@gmail.com}
\\
Department of Physics, University of Alberta\\
Edmonton, Alberta, Canada T6G 2E1
}
\date{(1994 July 4; slightly revised 2012 December 21,\\
Mayan Long Count Calendar 13.0.0.0.0, \\or day 1,872,000, end of the 13th b'ak'tun)}

\maketitle
\large
\begin{abstract}
\baselineskip 25pt

The observation that we are among the first $10^{11}$
or so humans reduces the prior probability that
we find ourselves in a species whose total lifetime number
of individuals is much higher, according to arguments of
Carter, Leslie, Nielsen, and Gott. However, if we instead
start with a prior probability that a history has
a total lifetime number which is very large,
without assuming that we are in such a history,
this more basic probability is not reduced by
the observation of how early in history we exist.
\end{abstract}
\normalsize
\pagebreak
\baselineskip 14.8 pt

John Leslie [1-16],
expanding upon largely unpublished
lectures by Brandon Carter \cite{C}, has argued that our observed
position in history increases, perhaps greatly, the probability that
the human race will soon end.  Variants of this argument have also
been discovered independently by Holgar Nielsen
\cite{N} and by J. Richard Gott III \cite {G}.

The idea is that if humans were to
continue at present or growing populations for more than a few
hundred additional years, it would be unlikely for us to have found
ourselves in the relatively small fraction alive by now.  On the
other
hand, if the human race were to end today (e.g., at the end of the Mayan 13th b'ak'tun when this paper is being revised), we would not be so
unusual, since about 5-10\% of all humans are alive today [20-22,19].
This possibility, or any doom within the next few hundred years,
does not make it nearly so unlikely for us to find ourselves alive
now.

Of course, what we are more interested in is the actual conditional
probability (often called a posterior probability)
that the human race will end soon, given that we are
here, rather than the reverse conditional probability
(often called a likelihood) that we are here, given that the human
race will end soon.  To calculate the former from the latter, we
need the prior probabilities that the human race will end
at various times in the future (or after various total numbers $N$
of people may have lived), and then we can apply Bayes' rule.
However, we do not have universally accepted prior probabilities,
so we cannot actually calculate universally accepted
posterior probabilities.

Without agreeing on the prior probabilities for the various
possibilities
for the total number $N$ of humans in an entire human history,
one can only use the observation of one's position
$N_0$ within human history to say how the prior probabilities
would need to be adjusted to give the posterior probabilities.
For example, suppose the prior probability for an observer to be
somewhere in a human race containing a total of $N$ people
thoughout its history were $P(N)$.  For simplicity, we shall
consider only probabilities conditional upon the existence of
a human race, so $N\geq 1$ and $P(N)$ is normalized to unity
when summed from $N=1$ to $\infty$.

The na\"{\i}ve result of incorporating the observation of the
position $N_0$ of the observer
would be the probability of $N$, given that $N$ is at least $N_0$,
	\begin{equation}
	P(N|N\geq N_0)=\theta(N-N_0)P(N)/\sum_{n=N_0}^{\infty}{P(n)},
	\label{eq:1}
	\end{equation}
where  $\theta(N-N_0)$ is 1 if $N\geq N_0$ and is 0 otherwise.
$P(N|N\geq N_0)$ is simply the prior probability $P(N)$ truncated
for all impossible situations $N<N_0$ and renormalized.

However, the point of the doomsday argument is to use the fact
that, within the range of the necessary condition $N\geq N_0$,
the larger $N$ is, the smaller is the likelihood or conditional
probability that the observer is the $N_0$th person within such
a history.  Here one adopts the following simplifying hypothesis:

\newpage

{\bf Assumption 1}

In a given history of length $N$, the probability is equal
for an observer to be in any of the $N$ possible positions (his or
her
possible birth order in the history, the value of $N_0$). That is,
the normalized conditional probability for the observer to have
position $N_0$, given the length $N$, is
	\begin{equation}
	P(N_0|N)=\theta(N-N_0)/N.
	\label{eq:2}
	\end{equation}

This is the assumption that the observer's birth order is purely
random
in the history.  It would follow from the assumption that the
observer is
a purely random human in the total history of $N$ people, but
it is weaker in that it does not require that any of the other
characteristics of the observer be random.  It is admittedly
unrealistic if the observer has special characteristics that
are correlated with his or her position.  For example, if the
observer
refers to a person reading this paper (e.g., you), he or she probably
has a knowledge of the English language that would make it
less likely for him or her to be the first human than to be a much
later
human.  Nevertheless, Leslie's argument against this type
of objection, (IId) in \cite{L4}, essentially the argument that
Carter
is right to select an observer by characteristics that are not
correlated with birth order, seems plausible, and is perhaps even
necessary in order to be able to use anthropic reasoning at all,
so we shall hereafter simply accept Assumption 1,
except when otherwise stated.

{}From Assumption 1 it follows that the joint probability that the
total history has $N$ total people and that the observer
is the $N_0$th one is
	\begin{equation}
	P(N,N_0)=P(N)P(N_0|N)=\theta(N-N_0)P(N)/N.
	\label{eq:3}
	\end{equation}
{}From this joint probability one can readily compute the
marginal probability distribution for $N_0$ alone as
	\begin{equation}
	P(N_0)=\sum_{N}^{}{P(N,N_0)}
	=\sum_{N=N_0}^{\infty}\frac{P(N)}{N}.
	\label{eq:4}
	\end{equation}
Then the posterior or conditional probability for $N$,
given an observed value of $N_0$, is, by Bayes' rule,
	\begin{equation}
	P(N|N_0)=\frac{P(N,N_0)}{P(N_0)}
	=\theta(N-N_0)N^{-1}P(N)/\sum_{n=N_0}^{\infty}{n^{-1}P(n)}.
	\label{eq:5}
	\end{equation}

A comparison with Eq. (\ref{eq:1}) shows that the true
posterior probability $P(N|N_0)$ has the same form as the
na\"{\i}ve probability $P(N|N\geq N_0)$, except with $P(N)$
replaced by $N^{-1}P(N)$, or by a normalized
	\begin{equation}
	\widetilde{P}(N)=
	N^{-1}P(N)\theta(N-1)/\sum_{n=1}^{\infty}{n^{-1}P(n)},
	\label{eq:6}
	\end{equation}
which is weighted toward smaller values of $N$ than $P(N)$ is.
For example, the na\"{\i}ve expectation value of $N$,
simply given that $N\geq N_0$, is
	\begin{equation}
	E(N|N\geq N_0)\equiv\sum_{N}^{}{N P(N|N\geq N_0)}
	=\left( \sum_{N=N_0}^{\infty}{N P(N)} \right)/
	\left( \sum_{N=N_0}^{\infty}{P(N)} \right),
	\label{eq:7}
	\end{equation}
whereas the true posterior expectation value of $N$,
given that one is the $N_0$th person, is
	\begin{equation}
	E(N|N_0)\equiv\sum_{N}^{}{N P(N|N_0)}
	=\left( \sum_{N=N_0}^{\infty}{P(N)} \right)/
	\left( \sum_{N=N_0}^{\infty}{\frac{P(N)}{N}} \right).
	\label{eq:8}
	\end{equation}
Assuming that both of these expectation values
are defined and that $P(N)$ is nonzero for more than
one value of $N\geq N_0$, one can readily show
from the Cauchy-Schwarz inequality that
	\begin{equation}
	E(N|N_0)<E(N|N\geq N_0).
	\label{eq:9}
	\end{equation}
Thus the true posterior expectation value of the
total number of humans is lower than the na\"{\i}ve
expectation value from the prior probability distribution
$P(N)$ for histories containing the observer.

To illustrate these various quantities, suppose that
$P(N)$ has a power-law dependence on $N$ for
$N$ between 1 and some integer $L$, which we shall
take to be much larger than $N_0$, which we shall also
take to be large, as the historical record indicates.
(One way to get a power-law probability distribution
is to assume that the population grows exponentially
with time at one rate until it is suddenly destroyed
by a disaster that occur randomly in time at some
other expected rate.  However, we are simply using
a power law for illustrative purposes, without implying
that we believe the true distribution has this form.)
Then
	\begin{equation}
	P(N)=\theta(N-1)\theta(L-N)N^{-s}/
	\sum_{n=1}^{L}{n^{-s}},
	\label{eq:10}
	\end{equation}
	\begin{eqnarray}
	P(N|N\geq N_0)&=&\theta(N-N_0)\theta(L-N)
	N^{-s}/\sum_{N=N_0}^{L}{N^{-s}}\nonumber\\
	&\simeq&(s-1)\theta(N-N_0)\theta(L-N)
	N^{-s}/(N_0^{1-s}-L^{1-s}),
	\label{eq:11}
	\end{eqnarray}
	\begin{equation}
	P(N,N_0)=\theta(N-N_0)\theta(L-N)
	N^{-s-1}/\sum_{n=1}^{L}{n^{-s}},
	\label{eq:12}
	\end{equation}
	\begin{eqnarray}
	P(N|N_0)&=&\theta(N-N_0)\theta(L-N)
	N^{-s-1}/\sum_{n=N_0}^{L}{n^{-s-1}}\nonumber\\
	&\simeq&s\theta(N-N_0)\theta(L-N)
	N^{-s-1}/(N_0^{-s}-L^{-s}),
	\label{eq:13}
	\end{eqnarray}
	\begin{equation}
	E(N|N\geq N_0)
	=\left( \sum_{N=N_0}^{L}{N^{1-s}} \right)/
	\left( \sum_{N=N_0}^{L}{N^{-s}} \right)
	\simeq\left( {s-1 \over s-2} \right)
	{N_0^{2-s}-L^{2-s} \over N_0^{1-s}-L^{1-s}},
	\label{eq:14}
	\end{equation}
	\begin{equation}
	E(N|N_0)
	=\left( \sum_{N=N_0}^{L}{N^{-s}} \right)/
	\left( \sum_{N=N_0}^{L}{N^{-s-1}} \right)
	\simeq\left( {s \over s-1} \right)
	{N_0^{1-s}-L^{1-s} \over N_0^{-s}-L^{-s}}.
	\label{eq:15}
	\end{equation}

For example, if $s>2$, the case in which both of
these expectation values are defined even in the
limit $L\rightarrow\infty$, they are both of order
$N_0$ (unless $s-2$ is very small), with the ratio
	\begin{equation}
	{E(N|N_0) \over E(N\geq N_0)}
	\simeq 1 - {1 \over {(s-1)^{2}}}.
	\label{eq:16}
	\end{equation}
This would be the case in which one would expect
doom within the next few hundred years, whether
one used the na\"{\i}ve probability $P(N\geq N_0)$
or the better posterior probability $P(N|N_0)$.
The latter would only shorten the time to the
expected doom by a factor of order unity
(again, unless $s-2$ is very small).

If $1<s<2, E(N|N_0)$ remains about $sN_0/(s-1)$,
but $E(N|N\leq N_0)$ is about $(s-1)N_0^s L^{s-1}/(2-s)$
and hence is much larger for $L\gg N_0$.  If $0<s<1$,
$E(N|N_0)$ is about $sN_0^sL^{1-s}/(1-s)$, but
$E(N|N\leq N_0)$ is roughly $(1-s)L/(2-s)$, again much
larger.  However, if $s<0$, both $E(N|N_0)$ and
$E(N|N\leq N_0)$ are of order $L$, with their ratio again
being given by Eq. (\ref{eq:16}).  Therefore, for a
power-law prior probability distribution $P(N)$ up to
$L\gg N_0$, with exponent $-s$, only for $0{\
\lower-1.2pt\vbox{\hbox{\rlap{$<$}\lower5pt\vbox{\hbox{$\sim$}}}}\ }
s
{\ \lower-1.2pt\vbox{\hbox{\rlap{$<$}\lower5pt\vbox{\hbox{$\sim$}}}}\
}
2$ is the actual posterior expectation value for $N$ significantly
lower than the na\"{\i}ve expectation value.  Only for $s{\
\lower-1.2pt\vbox{\hbox{\rlap{$>$}\lower5pt\vbox{\hbox{$\sim$}}}}\ }
1$ is $E(N|N\leq N_0)$ of order $N_0$, but for most of this possible
range of $s$, i.e., for $s{\
\lower-1.2pt\vbox{\hbox{\rlap{$>$}\lower5pt\vbox{\hbox{$\sim$}}}}\ }
2$, the na\"{\i}ve estimate $E(N|N\leq N_0)$ is also of order $N_0$.
That is, if $P(N)$ has the power-law form (\ref{eq:10}) with
$L\gg N_0$, only for $1{\
\lower-1.2pt\vbox{\hbox{\rlap{$<$}\lower5pt\vbox{\hbox{$\sim$}}}}\ }
s
{\ \lower-1.2pt\vbox{\hbox{\rlap{$<$}\lower5pt\vbox{\hbox{$\sim$}}}}\
}
2$ is the doomsday argument needed to conclude that doom is
expected to loom soon.

Incidentally, Gott's analysis \cite {G} initially looks quite
defective in avoiding mention of $P(N)$ altogether and in
conflating $P(N|N_0)$ with $P(N_0|N)$ \cite{Bu},
but Gott later explained \cite {G94} that he was adopting
a specific ``appropriate vague Bayesian prior'' analogous
to assuming that $P(N)$ has the power-law
distribution (\ref{eq:10}) with $s=1$ and with the limit
$L\rightarrow\infty$.  Although in this limit $P(N)$ is not
normalizable except to zero, Eq. (\ref{eq:13}) does give a
normalizable $P(N|N_0)$, though one with an infinite expectation
value for $N$.  For example, for $L$ finite but much larger than
$N_0$, $E(N|N_0)\simeq N_0\ln{(L/N_0)}$, which grows
indefinitely with $L$, though at a much slower rate than
$E(N|N\geq N_0)\simeq L/\ln{(L/N_0)}.$

We are not claiming that such an assumption for $P(N)$
is necessarily unreasonable, and indeed one
can advance some reasons for preferring it if all that is known
about $N$ is that it is positive \cite {G94,J} (though even with that
unrealistic assumption it would seem more natural to apply it to
the probability distribution $P_0(N)$ to be defined below,
rather than to $P(N)$).  However, it is not obvious to us that this
assumption for $P(N)$ follows from ``only the assumption that
you are a random intelligent observer'' \cite {G}, unless one
implicitly defines the latter assumption to mean the former.

Thus if one starts with a prior probability $P(N)$ for the observer
to exist within a human history with a total of $N$ people, the
doomsday
argument weights the posterior probability distribution $P(N|N_0)$
toward lower values of $N$ than the na\"{\i}ve estimate
$P(N|N\leq N_0)$, but for this to have a significant effect,
$P(N)$ must have a rather restricted form.

However, the main point of the present paper is that instead
of starting with the probability $P(N)$ for a history containing
$N$ people that includes the observer in question, it would
be more natural to start with a probability $P_0(N)$ for
a history to contain $N$ people, without requiring
the observer's existence within it.  Certainly $P_0(N)$ would be more
basic and easier to calculate if one could ever get a theory giving
the probabilities for various human histories.

Again we shall only consider $N\geq 1$ and normalize $P_0(N)$
over such values.  That is, $P_0(N)$ is actually the conditional
probability that a human history has $N$ people,
given the condition that a history of at least one person exists.
This is obtained by dropping the possibility of no human history
(since this possibility is irrelevant to our present arguments),
and renormalizing the remaining probabilities we do consider.
Thus we actually use what we might write more precisely
as $p_0(N|N\geq 1)$, but we shall simply call this
conditional probability $P_0(N)$ for short.  However, the crucial
point is that in $P_0(N)$, we do not require the condition that
the observer in question be included in the history.

(Incidentally, the subscript $0$ on $P_0(N)$ here is intended
to give the connotation of a more basic probability distribution
than $P(N)$.  It is not intended to have the same
connotation as the subscript 0 on $N_0$, where it denotes
the observer in the history.  The latter usage has
a roughly similar connotation to the subscript 0 on the $t_0$
used in cosmology to denote the present age of the universe,
if the observer is taken to be a person at the
present point in history, for example, you.)

If one does start with $P_0(N)$ rather than $P(N)$, the na\"{\i}ve
result of incorporating the observation that the observer
is the $N_0$th person would be
	\begin{equation}
	P_0(N|N\geq N_0)
	=\theta(N-N_0)P_0(N)/\sum_{n=N_0}^{\infty}{P_0(n)}.
	\label{eq:17}
	\end{equation}
We would now like to compare this na\"{\i}ve result with the
result $P(N|N_0)$ of using Bayes' rule and the doomsday
argument.  Since that procedure led to Eq. (\ref{eq:5}), we can
continue to use it once we calculate $P(N)$ in terms of $P_0(N)$.

In the same spirit in which we previously adopted Assumption 1,
it is now simplest to assume that for two different histories of
equal probability (say with $N_1$ and $N_2$ people respectively),
the observer has an equal probability to be any of the $N_1+N_2$
people in both of these histories.  Then even though in this case
$P_0(N_1)=P_0(N_2)$, the probability that the history
contains the observer would be proportional to the
number of people in the history, $P(N_1)/P(N_2)=N_1/N_2$.
Extending this reasoning to histories with different existence
probabilities $P_0(N)$ leads to the following hypothesis:

{\bf Assumption 2}

The probability for the observer to exist somewhere in a history of
length $N$ is proportional to the probability for
that history and to the number of people in that history.
That is, the normalized probability for a history containing
the observer is
	\begin{equation}
	P(N)=NP_0(N)/\sum_{n=1}^{\infty}{nP_0(n)}
	=NP_0(N)/E_0,
	\label{eq:18}
	\end{equation}
where
	\begin{equation}
	E_0=\sum_{n=1}^{\infty}{nP_0(n)}
	\label{eq:19}
	\end{equation}
is the prior expected length of a human history, assuming
that one of some positive length $n\geq 1$ does exist.

(Of course, if the observer's detailed individual
characteristics were considered, such as a knowledge of an
estimate for $N_0$, the probabilities for one to be at different
positions in various histories of equal existence probabilities
would not be equal,
whether for a single history containing $N$ people, as discussed
above, or whether for two histories, containing $N_1$ and $N_2$
respectively.  Then the doomsday argument would lose its simple
applicability, but, as we discussed above for Assumption 1, we shall
assume that the characteristics defining the observer are
uncorrelated with $N_0$ and with $N$.)

Inserting Eq. (\ref{eq:19}) into Eq. (\ref{eq:3}), we find that the
joint probability for the history to have length $N$ and for the
individual to be the $N_0$th person in the history is
	\begin{equation}
	P(N,N_0)=\theta(N-N_0)P_0(N)/\sum_{n=1}^{\infty}{nP_0(n)}
	=\theta(N-N_0)P_0(N)E_0^{-1},
	\label{eq:20}
	\end{equation}
where $E_0^{-1}$ may be considered as the prior probability
for one to be the $N_0$th person in a history at least as long
as $N_0$, a constant under Assumptions 1 and 2.

Once we use Eq. (\ref{eq:20}) in Eq. (\ref{eq:5}), we find that
it gives exactly the same result as Eq. (\ref{eq:17}), the na\"{\i}ve
consequence of $P_0(N)$.  That is,
	\begin{equation}
	P(N|N_0)=P_0(N|N\geq N_0),
	\label{eq:21}
	\end{equation}
as the weighting toward smaller $N$ that the doomsday
argument provides is precisely canceled by the greater
probability of finding the observer within the larger of two sets of
people whose prior existence is equally probable.  In other words,
the doomsday argument has no effect at all if, instead of
starting with $P(N)$, we start with the more natural prior
probability $P_0(N)$ for a history containing
$N$ humans.  Incidentally, one can readily see that $P_0(N)$
is the same as $\widetilde{P}(N)$ defined by Eq. (\ref{eq:6}).

As a simple analogue to illustrate our main point, consider
two north-south roads with $N_1$ and $N_2$ houses along
each respectively, say with $N_1\ll N_2$.  If we choose
randomly between the two roads, $P_0(N_1)=P_0(N_2)=1/2$.
However, if we choose randomly between the $N_1+N_2$
houses, each has a probability $1/(N_1+N_2)$, so the
probability of having the first road along a random house is
$P(N_1)=N_1/(N_1+N_2)\ll P(N_2)=N_2/(N_1+N_2)$.

Now suppose we observe that the house is the $N_0$th
from the north end, with $N_0\leq N_1\ll N_2$.  Starting
from the unequal probabilities $P(N_1)$ and $P(N_2)$,
we would get the same two unequal numbers for the
na\"{\i}ve probabilities $P(N_1|N_1\geq N_0)$ and
$P(N_2|N_2\geq N_0)$, but these are obviously not
the true probabilities, since the $N_0$th house from
the north end is certainly not a house chosen randomly
from the $N_1+N_2$ houses.

For our example, the analogue of the doomsday argument
corrects for this error.  The likelihood for a random house
to be $N_0$th from the north end is $P(N_0|N_1)=1/N_1$
if it is on the first road and $P(N_0|N_2)=1/N_2$ if it is on
the second.  Although $P(N_0|N_1)\gg P(N_0|N_2)$, that
does not imply $P(N_1|N_0)\gg P(N_2|N_0)$ as Gott's
analysis \cite {G} implicitly assumes, but it does combine
with $P(N_1)$ and $P(N_2)$ to give equal joint probabilities
	\begin{equation}
	P(N_1,N_2)\equiv P(N_0|N_1)P(N_1)=P(N_2,N_0)\equiv
	P(N_0|N_2)P(N_2)=1/(N_1+N_2)
	\label{eq:22}
	\end{equation}
and equal posterior probabilities
	\begin{equation}
	P(N_1|N_0)=P(N_2|N_0)=1/2.
	\label{eq:23}
	\end{equation}
However, these equal posterior probabilities are obviously
the same as what Eq. (\ref{eq:17}) would give directly for
 $P_0(N_1|N_1\geq N_0)$ and $P_0(N_2|N_2\geq N_0)$
without having to bother with doomsday-like arguments.

In other words, if the two roads, with their $N_1$ and $N_2$
respective houses, have equal prior probability, these
probabilities are not affected by the observation that a
house chosen at random from all of the houses is at the
$N_0$th position (assuming $N_0\leq N_1$ and $N_0\leq N_2$).
Each road has an equal number of houses at the $N_0$th
position (namely, one), so the observation of this position
does not affect the probabilities for the two roads.

As a weaker alternative to Assumptions 1 and 2, one could
start with the following hypothesis instead:

{\bf Assumption 3}

The length $N$ of a history and the observer's position $N_0$
in a history are independent random variables,
except for the trivial restriction $1\leq N_0\leq N$.
That is, the joint probability has the essentially product form
	\begin{equation}
	P(N,N_0)=P_0(N)p_0(N_0)\theta(N-N_0),
	\label{eq:24}
	\end{equation}
where $P_0(N)$ may be considered to be the prior
existence probability for a history of length $N$,
and where $p_0(N_0)$ (using a lower case $p$ for distinction)
may be considered to be the prior probability
for the observer to be the $N_0$th person in a history
at least as long as $N_0$.

A comparison of Eqs. (\ref{eq:20}) and (\ref{eq:24}) shows
that Assumptions 1 and 2 give the special case
$p_0(N_0)=E_0^{-1}$, a constant, so Assumption 3 is
indeed more general.  However, even it is ultimately unrealistic
if one includes any special characteristics in the definition
of the observer in question.  For example, if that observer
is a person who knows of the existence of nuclear weapons
of mass destruction, one might reasonably expect that
not only does that characteristic makes $p_0(N_0)$ larger
for $N_0$ near $10^{11}$, say, than for $N_0$ near unity, but
also it might reduce the probability $P(N,N_0)$ for $N\gg N_0$
below what the product form (\ref{eq:24}) would give,
due to the greater probability of the ending of the human
history by the weapons known to the observer.
Nevertheless, in the same spirit in which we considered
Assumptions 1 and 2, we shall consider Assumption 3
here for the sake of argument.

	From Eq. (\ref{eq:24}) one can readily calculate that
the marginal distributions for $N$ and $N_0$ are
	\begin{equation}
	P(N)=P_0(N)\sum_{N_0=1}^{N}{p_0(N_0)},
	\label{eq:25}
	\end{equation}
	\begin{equation}
	P(N_0)=p_0(N_0)\sum_{n=N_0}^{\infty}{P_0(N)}.
	\label{eq:26}
	\end{equation}
Therefore, instead of Eq. (\ref{eq:2}), we get
	\begin{equation}
	P(N_0|N)=P(N,N_0)/P(N)
	=p_0(N_0)\theta(N-N_0)/\sum_{n_0}^{N}{p_0(N_0)}.
	\label{eq:27}
	\end{equation}
However, we continue to get Eq. (\ref{eq:21}),
$P(N|N_0)=P_0(N|N\geq N_0)$, so even with the
weaker Assumption 3, the doomsday argument is
not needed if we start with the $P_0(N)$ in Eq. (\ref{eq:24})
and its na\"{\i}ve consequence $P_0(N|N\geq N_0)$
defined by Eq. (\ref{eq:17}).

This says that we cannot obtain any information about
the length $N$ of human history from the knowledge
of an observer's position $N_0$ in history (except for the trivial
restriction $N\geq N_0$).  This is an obvious consequence
of Assumption 3, which says that, except for the trivial
restriction, the length and the position are independent.

Of course, the advocate of the doomsday argument
can still point out that if one instead started with the
marginal distribution $P(N)$ given in Eq. (\ref{eq:25}),
and then used Eq. (\ref{eq:1}) to define the na\"{\i}ve
$P(N|N\geq N_0)$, that would not be the same as
$P(N|N_0)$, and so doomsday arguments would be
needed to correct this alternate na\"{\i}ve result.
Because of that fact, we cannot claim to have proved
that the doomsday argument is absolutely wrong.  We
only claim that it is unnecessary if one starts with
$P_0(N)$ instead of $P(N)$, and that both as a prior
probability and as a component of
Eq. (\ref{eq:25}), $P_0(N)$ appears to be more
basic than $P(N)$.

After formulating the present objection against the
necessity of using doomsday arguments, we found that
Leslie \cite{L4,L13} had already discussed it, though
he was unconvinced by this objection.  He gives one form
of the objection as (IIIa) in \cite{L4}:
``The larger our race is in its temporal entirety, the more
opportunities there are of being born into it.  This
counterbalances the greater unlikelihood of being born early."
Then he responds, ``This seems to me false.  We are not here
dealing with some ordinary lottery where we would have existed
in numbers that remained constant whether or not we had got
tickets, so that possessing a ticket could itself readily suggest
that many were sold or thrown to the crowd; for it seems wrong
to treat ourselves as if we were once immaterial souls harbouring
hopes of becoming embodied, hopes that would have been greater,
the greater the number of bodies to be created. \ldots"
However, this objection shifts the doomsday argument
from its simple setting towards the question of choosing
in a sophisticated way $P(N)$ or $P_0(N)$---a separate problem
which we shall not discuss here, though one of us may discuss
it in a future publication.

We do not say that we believe that human history
will necessarily continue long into the future.  There are
many reasons, such as the history of perished species,
overpopulation, dwindling resources, pollution, disease,
technological capabilities for destruction, the aggressive
nature of humans, and religious revelation, to suggest that
human history , at least as we know it, could well end rather
soon.  For example, if $P_0(N)\propto N^{\alpha}$ with
$\alpha=-s-1<-2$, then given one's position $N_0$,
the expected total number of humans, $E(N|N_0)$,
is of the same order as $N_0$, and if $\alpha<-1$,
most of the total probability is for $N\sim N_0$.
We are also not saying that the end of human
history as we know it is necessarily best described
as a gloomy ``doom," a semantic objection that
Carter also has \cite{CL} with the name Leslie says
Frank Tipler gave to the argument \cite{LP},
but it was irresistible for us
to be able to include four double o's in our title.
However, what we have shown is that the
Carter-Leslie-Nielsen-Gott doomsday argument,
at the level at which we discuss it, is
inconclusive in predicting how soon there may be ``doom."

Our consideration of this topic has been especially
motivated by long discussions with John Leslie, for
which we are grateful.  After our first draft was written,
we have benefited from lengthy responses from Brandon
Carter, Leslie, and J. R. Gott III, who continue to disagree
with us but have not convinced us to abandon our objection
to the doomsday argument.  We have also benefited from
discussions with Geoffrey Hayward, Werner Israel, and
others at a CIAR conference in Lake Louise and at the
University of Alberta, where some of these ideas have
been presented in seminars.  This work has been supported
in part by the Natural Sciences and Engineering Research
Council of Canada.  We have also been motivated by the Mayans to submit this paper to a journal on this particular auspicious day.

\newpage

\end{document}